\documentclass[11pt,a4paper]{article}

\pdfoutput=1

\usepackage{graphicx}
\usepackage{comment}
\usepackage{epstopdf}
\usepackage{jcappub}
\usepackage{amsmath,amsthm,latexsym,amssymb,amsfonts}
\usepackage{fontenc,layout}
\usepackage{hyperref}
\usepackage{psfrag}
\usepackage[usenames,dvipsnames]{xcolor}

\usepackage[utf8]{inputenc}
\usepackage{indentfirst}

\numberwithin{equation}{section}

\title{Gravitational wave signals and cosmological consequences of gravitational reheating
}

\author[a,b]{Micha{\l} Artymowski}
\author[a]{Olga Czerwi\'nska}
\author[a]{Zygmunt Lalak}
\author[a,c]{Marek Lewicki}
\affiliation[a]{Institute of Theoretical Physics, Faculty of Physics, University of Warsaw, ul. Pasteura 5, 02-093 Warsaw, Poland}
\affiliation[b]{Institute of Physics, Jagiellonian University, {\L}ojasiewicza 11, 30-348 Krak{\'o}w, Poland}
\affiliation[c]{Kings College London, Strand, London, WC2R 2LS, United Kingdom}
\emailAdd{Michal.Artymowski@fuw.edu.pl}
\emailAdd{Olga.Czerwinska@fuw.edu.pl}
\emailAdd{Marek.Lewicki@kcl.ac.uk}
\emailAdd{Zygmunt.Lalak@fuw.edu.pl}

\abstract{
Reheating after inflation can proceed even if the inflaton couples to Standard Model (SM) particles only gravitationally. However, particle production during the transition between de-Sitter expansion and a decelerating Universe is rather inefficient and the necessity to recover the visible Universe leads to a non-standard cosmological evolution initially dominated by remnants of the inflaton field. We remain agnostic to the specific dynamics of the inflaton field and discuss a generic scenario in which its remnants behave as a perfect fluid with a general barotropic parameter $w$. Using CMB and BBN constraints we derive the allowed range of inflationary scales. We also show that this scenario results in a characteristic primordial Gravitational Wave (GW) spectrum which gives hope for observation in upcoming runs of LIGO as well as in other planned experiments.
}


\begin{document}
\begin{flushright}
KCL-PH-TH/2017-59
\end{flushright}
\maketitle

\section{Introduction} \label{sec:Introduction} 

Cosmic inflation \cite{Lyth:1998xn,Starobinsky:1980te,Mazumdar:2010sa} is a well established theory describing the evolution of the very early Universe that exhibits strong consistency with the recent experimental data \cite{Ade:2015lrj,Array:2015xqh}. It predicts accelerated expansion of the scale factor, which solves many problems of the classical Big Bang cosmology \cite{Guth:1980zm}. Cosmic inflation dilutes any pre-existing matter and radiation and thus requires a reheating mechanism \cite{Felder:1998vq,Kofman:1997yn,Kofman:1994rk} to eventually result in the Universe dominated by radiation. This is usually realised by assuming couplings between the inflaton and matter fields. However, strong couplings may lead to non-trivial loop corrections to inflationary potential, which in principle may spoil its flatness. In consequence, this could spoil the predictions of inflation \cite{Gaillard:1995az}. 
\\*

This motivates us to study alternative models of reheating such as gravitational reheating in which production of particles after inflation proceeds only due to gravitational interactions \cite{Ford:1986sy,Kunimitsu:2012xx,Dimopoulos:2017zvq,Peebles:1998qn,deHaro:2017nui,AresteSalo:2017lkv}.
In particular the transition between de-Sitter evolution with the exponential growth of the scale factor and a decelerating Universe produces quantum modes of scalar fields, which at certain point may dominate the Universe and increase its temperature sufficiently. In this scenario the inflaton does not need to be coupled to any standard model degree of freedom and therefore it may be a part of a dark sector. In this work we denote an inflaton with only gravitational coupling to SM as a dark inflaton and our main purpose is to investigate possible phenomenological consequences of this scenario independently of the particular structure of the dark and inflationary sector. 
\\*

Another reason to study gravitational particle production as the mechanism for reheating is the uncertainty {in} the thermal history of the Universe. Due to said uncertainty one cannot calculate the exact moment of the horizon crossing of the pivot scale \cite{Liddle:2003as,Martin:2010kz}. Using the fact that dark inflation can be followed by domination of a perfect fluid with a barotropic parameter $w$ and then by the usual radiation and matter domination, we calculate the number of e-folds before the end of inflation at the pivot scale horizon crossing \cite{Ade:2015lrj} 
\begin{equation}
N_\star \simeq {67} - \log\left(\frac{k_\star}{a_0 H_0}\right) + \frac{1}{4}\log\left(\frac{V_{hor}}{M_p^4}\right) + \frac{1}{4} \log\left(\frac{V_{hor}}{\rho_{end}}\right) + \frac{1-3 w}{12(1+ w)}\log\left(\frac{\rho_{th}}{\rho_{end}}\right) \, , \label{eq:Nstar}
\end{equation}
where $k_\star$ is the pivot scale, $a_0 H_0$ is the inverse of the comoving Hubble radius today, $V_{hor}$ is the value of the inflaton potential at the horizon crossing, $\rho_{end}$ is the scale of the end of inflation, $\rho_{th}$ is the energy scale at which radiation starts to dominate, while $w$ defines the equation of state between the last two.
\\*

The most stringent current experimental constraint on the temperature of reheating comes from the Big Bang Nucleosynthesis, which occurs at the MeV scale~\cite{Cyburt:2015mya}.
{Thus, the} reheating energy scale $\rho_{th}$ can in principle take any value from $\rho_{th} \sim \rho_{end}$ (an instant reheating scenario) to $\rho_{th} \sim \text{MeV}^4$.
This ambiguity significantly affects $N_\star$ and therefore also the predictions of inflationary models.
In the dark inflationary scenario we obtain a reheating temperature as a precise function of inflationary parameters, such as scale of inflation and the {post-inflationary equation of state. In addition, most of the reheating mechanisms require the existence of additional couplings between the inflaton and matter fields. Gravitational particle production always occurs at the end of inflation, regardless of the form of the inflationary potential. It does not require any additional interactions between fields, therefore dark inflation} can decrease the amount of new physics needed in order to explain the matter content of the present Universe.
\\*

As a {component}  of the dark sector of the Universe, the inflaton can be related to dark matter (DM) and dark energy (DE). This could serve as an additional motivation fort this work. 
\\*

Throughout the paper we will use the convention $8\pi G = M_{p}^{-2}$, where $M_{p} = 2.435\times 10^{18}$ GeV is the reduced Planck mass.

\section{Gravitational particle production} \label{sec:reheating}

In this section we proceed to investigate the gravitational particle production during the transition era between the de-Sitter expansion, which is a good approximation of the cosmic inflation era, and a decelerating Universe. Let us consider the evolution of the inflaton and a scale factor as a function of a conformal time $\eta$ defined by
\begin{equation}
a(\eta)d \eta = dt \, .
\end{equation}

As shown in the Ref \cite{Ford:1986sy}, the energy density of radiation generated by gravitational particle production for the inflaton minimally coupled to gravity is equal to 
\begin{equation}
\rho_r = \frac{H_{\rm inf}^4}{128\pi^2}\left(\frac{a_{end}}{a}\right)^4 I \, , \label{eq:reh}
\end{equation}
where $H_{\rm inf}$ is the value of the Hubble parameter at the plateau and $a_{end}$ is a value of the scale factor at the and of inflation.  The integral $I$ is defined by
\begin{eqnarray}
&I = -\int \limits_{-\infty}^{x} dx_1 \int \limits_{-\infty}^{x} dx_2 \log(|x_1-x_2|)\frac{d\tilde{V}(x_1)}{dx_1} \frac{d\tilde{V}(x_2)}{dx_2} \, ,\\
&\tilde{V}(x) = \frac{f_{xx}f - \frac{1}{2}f_{x}^2}{f^2} \, ,\\
&f(H_{\rm inf}\eta) = a^2(\eta) \, ,
\end{eqnarray}
where $\tilde{V}$ is a rescaled Ricci scalar, $x = H_{\rm inf}a$ and $f_x = \frac{df}{dx}$. The upper limit for the integration of $I$ corresponds to the moment when $\tilde{V} < 1$. 
 
Gravitational particle production rate can be estimated analytically by investigating the transition between de-Sitter Universe and a decelerating solution, as in Refs. \cite{Ford:1986sy,Kunimitsu:2012xx}. Here we want to generalise their analysis assuming that the decelerating Universe is filled with any perfect fluid with a constant barotropic parameter $w$. We can describe the evolution of $g(x)$ in the following way
\begin{equation}
f(x) = \begin{cases} \frac{1}{x^2} \quad \quad \text{$x < -1$, de Sitter} \\
a_0 + a_1 x + a_2 x^2 + a_3 x^3 \quad \quad \text{$-1 < x < x_0 -1$, transition} \\
b_0( b_1+  x)^{\frac{4}{3w+1}} \quad \quad \text{$x_0 - 1 < x$, general $w \neq -1/3$} 
\end{cases}
\end{equation}
where $x_0$ is the transition time between the de-Sitter and decelerating solutions. The $a_i$ and $b_i$ coefficients can be calculated using continuity conditions for $f(x)$, $f'(x)$ and $f''(x)$ at $x = -1$ and $x= x_0 - 1$. Assuming $x_0 \ll 1$ one finds
\begin{eqnarray}
&a_0 = \frac{1}{4} \left(29-8 w+3 w^2\right)-\frac{1+w}{2x_0} \, , \label{eq:a0} \\
&a_1 = \frac{3}{4} \left(\frac{47}{3}-8 w+3 w^2\right)-3\frac{1+w}{2x_0} \, , \label{eq:a1} \\
&a_2 = \frac{3}{4} \left(9-8 w+3 w^2\right)-3\frac{1+w}{2 x_0} \, ,  \label{eq:a2} \\
&a_3 = \frac{1}{4} \left(5-8 w+3 w^2\right)-\frac{1+w}{2 x_0} \, , \label{eq:a3} \\
&b_0 = \left(\frac{2}{1+3 w}\right)^{-\frac{4}{1+3 w}} \, ,  \label{eq:b0} \\
&b_1 = \frac{3 (1+w)}{1+3 w} \, . \label{eq:b1}
\end{eqnarray}
The biggest contribution to $I$ comes from integration around the transition time, i.e. for $x\in (x_0-1,-1)$. Therefore it can be estimated by
\begin{equation}
I \simeq 9 (w+1)^2 \log \left(\frac{1}{x_0}\right) \, .\label{eq:I}
\end{equation}
This result is consistent with~\cite{Ford:1986sy,Kunimitsu:2012xx}, but it does not fully apply to the $w=-1/3$ case, {when $f(x) = b_0 \exp(b_1 x - 1)$ for $x>x_0-1$ and the only change is that the continuity conditions imply that $b_0 = 1/2$ and $b_1 = 2$ up to $\mathcal{O}(x_0)$ terms.} One could also consider the transition between two de Sitter space-times with different values of $H_{\rm inf}$ - in such a case one finds $f = b_0/(x+b_1)^2$ for $x > x_0 - 1$. Surprisingly the $a_i$ coefficients satisfy Eq. (\ref{eq:a0}-\ref{eq:a3}) for $w = -1$, which means that there is no $x_0^{-1}$ term corresponding to $I \propto x_0^2$. In such a case particle production is strongly suppressed, since one assumes $x_0 \ll 1$.
\\*

Using Eq.~\eqref{eq:reh} the energy density of radiation produced at the end of inflation can be estimated as
\begin{equation}
\rho_r \simeq H_{\rm inf}^4\frac{9 N_{\rm eff} (1+w)^2}{128\pi^2} \left( \frac{a_{end}}{a} \right)^4 \log \left(\frac{1}{x_0}\right) \, \label{ro_r},
\end{equation}
where $N_{\rm eff}$ for an inflaton minimally coupled to gravity is the number of scalar species produced gravitationally\footnote{Fermions and vectors may be also produced during the transition between two gravitational vacua. Nevertheless, energy densities related to them are too small to significantly contribute to the reheating of the Universe.}. Non-minimal coupling of the inflaton to gravity results in an additional numerical coefficient in Eq. \eqref{ro_r}. For example, in a specific case of a scalar-tensor theory with a non-minimal coupling of the form $\xi \phi^2 R$ we would have $N_{\rm eff} = N (1-6\xi)^2$. Thus the gravitational particle production can be strongly amplified by big values of $\xi$ or strongly suppressed by the non-minimal coupling close to the conformal value. We want to remain as model-independent as possible so we will simply include a wide range of possible values of $N_{\rm eff}$ and show the consequences of each choice. The $\log (1/x_0)$ term should be of order of unity \cite{Ford:1986sy} and it will be neglected in the further part of this analysis. 
\\*

Since gravitational particle production is highly inefficient compared to most other mechanisms of reheating \cite{Kunimitsu:2012xx}, dark inflation leads to the Universe initially dominated by the inflaton field. For the evolution to result in a radiation-dominated Universe, which is necessary at least in the BBN era, we need the energy density of the inflaton redshifting faster than radiation after inflation. This means that dark inflation changes the thermal history of the Universe dramatically by introducing a long period of inflaton domination in the post-inflationary era. We will assume that the inflaton field may be treated as a perfect fluid with constant barotropic parameter $w$ and we will limit ourselves to $w \in [1/3,1]$.
\\*

There are many possible realisations of the dark sector leading to post-inflationary evolution with $w > 1/3$~\cite{Joyce:1997fc}. Possibly the simplest is just using an inflationary potential $V \propto \phi^{2n}$ with $n\geq 3$. After inflation the inflaton will oscillate around the minimum of the potential which leads to an equation of state \cite{Joyce:1997fc,Ford:1986sy} with
\begin{equation}
w = \frac{n-1}{n+1} \, .
\end{equation}
For $n \to \infty$ one finds $w \to 1$ which is also the result for an exponential potential or in fact any potential in which the kinetic energy dominates over the potential contribution. However, inflation generically requires a flat potential and for the above to be viable we need to work in the framework of $\alpha$-attractors \cite{Kallosh:2013yoa,Artymowski:2016pjz,Dimopoulos:2016yep} or scalar-tensor theories \cite{Kallosh:2014laa,Giudice:2014toa,Artymowski:2016ikw}. In both cases the potential takes form of a flat plateau for big absolute values of the inflaton field, which leads to inflationary models consistent with predictions of the Starobinsky $R^2$ inflation \cite{Starobinsky:1980te}.
\\*

{As shown in Refs. \cite{Watanabe:2007tf,Ema:2015dka,Ema:2016hlw} the oscillating scalar field may produce other scalar fields during its oscillation phase. This effect may be significant even if one assumes the lack of direct couplings between the oscillating inflaton and the subdominant fields. In such a case the mechanism of gravitational reheating may also help to sufficiently reheat the Universe before the BBN.}
\\*

Another possible realisation of the inflationary sector which gives post-inflationary evolution with $w>1/3$ are models, where the accelerated expansion is driven by non-canonical kinetic terms. Two examples of such theories, which recently have been investigated in the context of the gravitational particle production, are G-inflation \cite{Kobayashi:2010cm} and K-inflation \cite{ArmendarizPicon:1999rj,Garriga:1999vw,Helmer:2006tz,Kunimitsu:2012xx}. In both cases inflation ends with the massless scalar field domination, which corresponds to $w=1$. 
\\*

\begin{figure}[ttt]
\centering
\includegraphics[height=5.1cm]{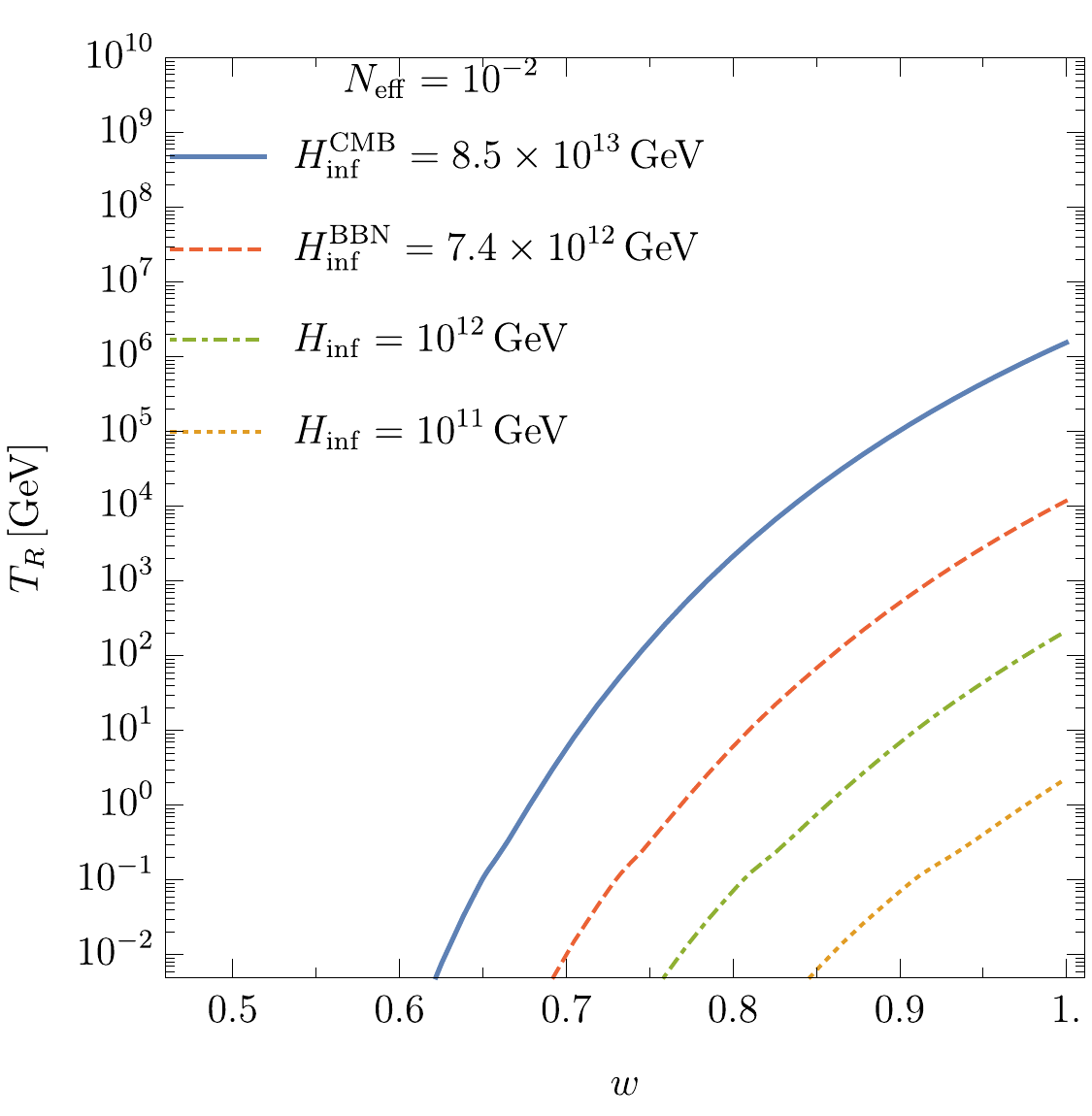}
\includegraphics[height=5.1cm]{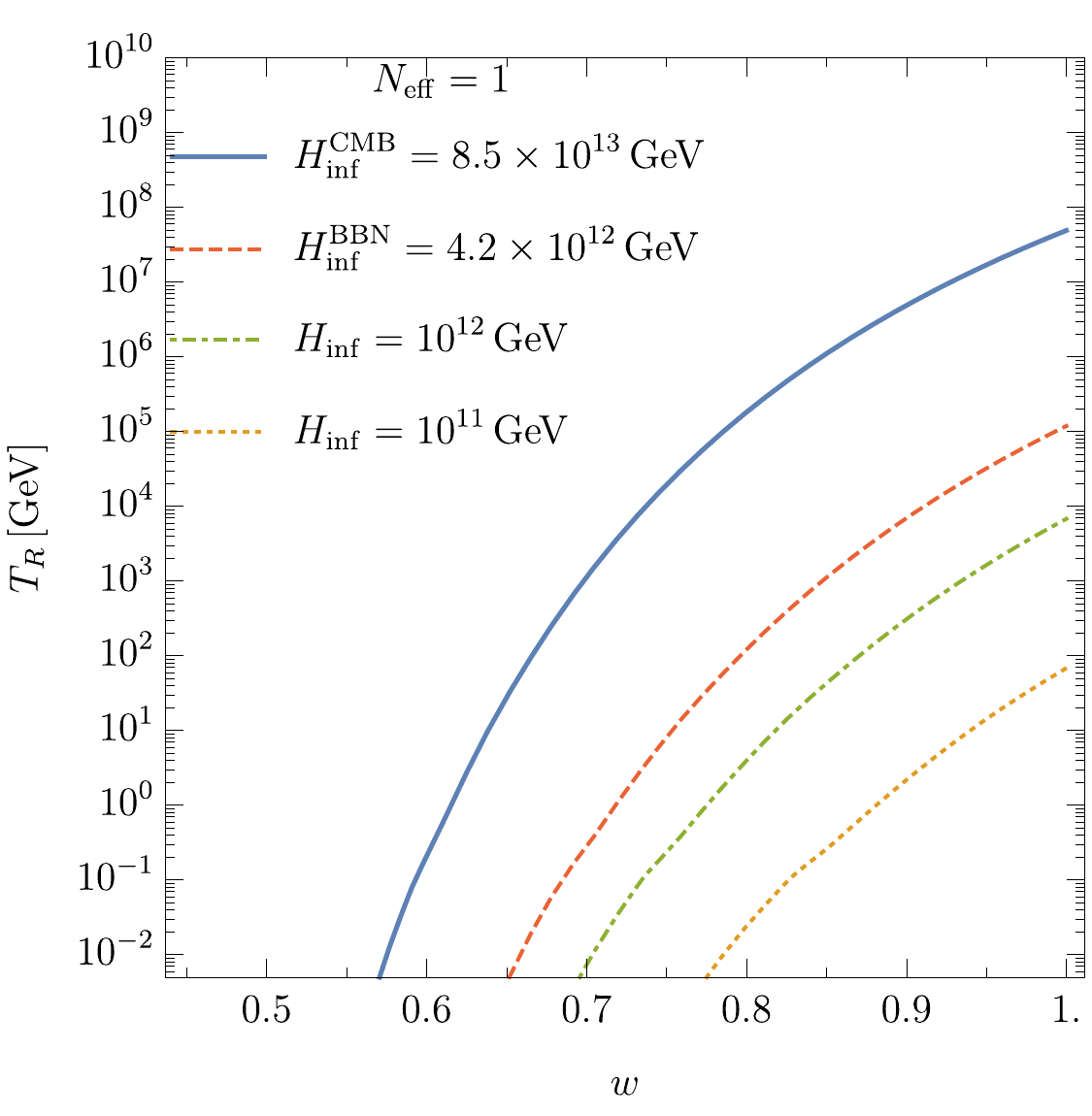}
\includegraphics[height=5.1cm]{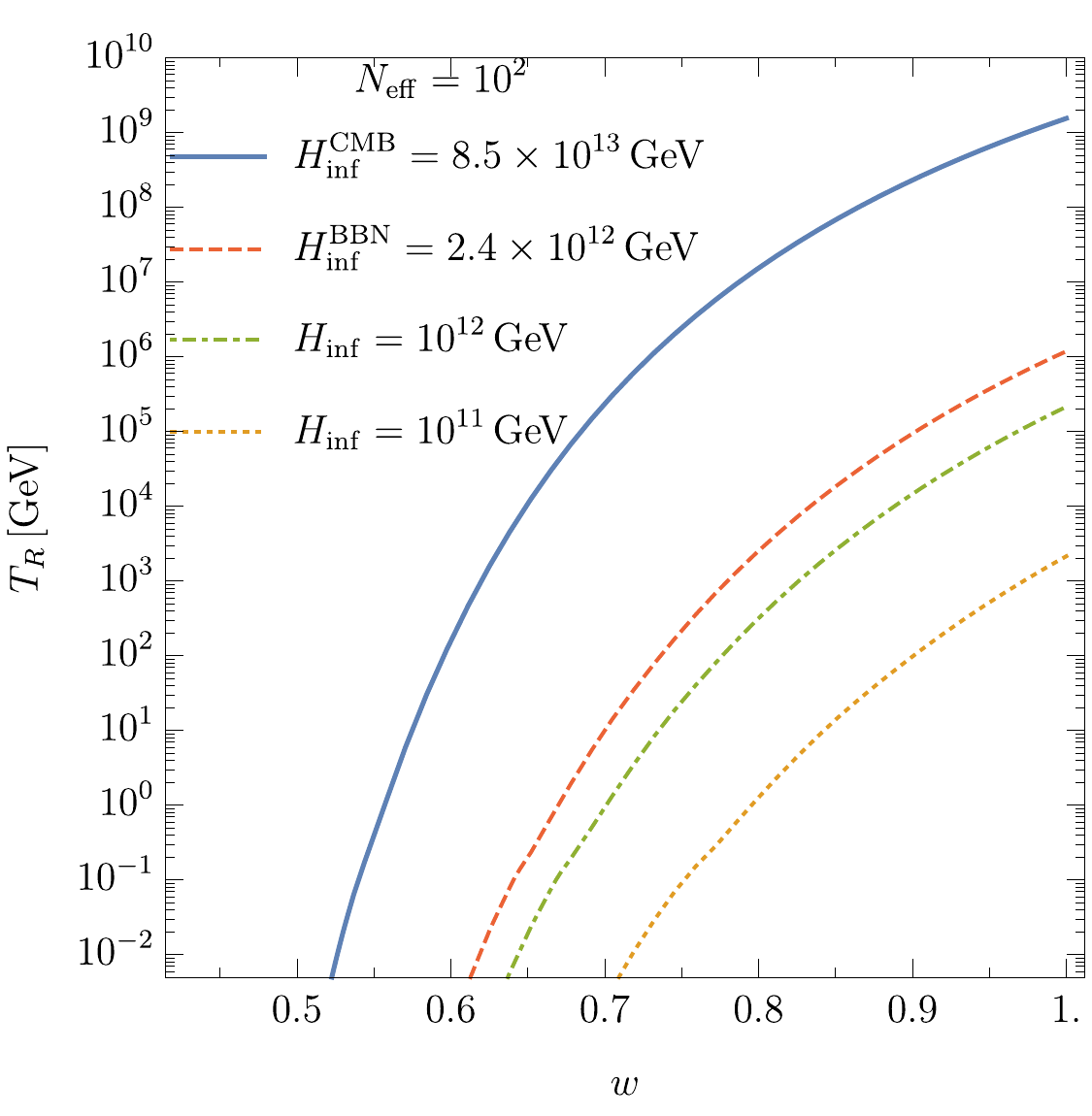}
 \caption{\label{fig:Treh}
Temperature $T_R${,} after which the usual radiation dominated history of the Universe resumes{,} as a function of $w$ for several inflationary scales $H_{\rm inf}$ with $N_{\rm eff} \in \{10^{-2}, 1, 100\}$ from left to right.}
 \end{figure}

Note that the moment when radiation is produced is not equivalent to reheating, which we define as the moment when radiation comes to dominate the energy density of the Universe. Then the reheating temperature $T_R$ is set by the condition $\rho_r = \rho_{\phi}$, which gives
\begin{equation}
\rho_{\phi} = 3 H_{\rm inf}^2  M_p^2 \left(  \frac{a_{end}}{a} \right)^{3(1+w)}\, ,\label{ro_fi}
\end{equation}
where $a_{end}$ is the scale factor at the end of inflation and $\rho_{\phi}$ is the energy density of the inflaton field. Using \eqref{ro_r} and \eqref{ro_fi} we find the scale factor at the moment of reheating
\begin{equation} \label{eq:aR}
a_R =  a_{end} \left(\frac{128 \pi^2  M_p^2}{3 N_{\rm eff} (1+w)^2 H_{\rm inf}^2}\right)^{\frac{1}{3w-1}} \, 
\end{equation}
 and the radiation energy density at that time
\begin{equation}\label{eq:rhoR}
\rho_R \equiv \rho_r (a_R) = 3 H_{\rm inf}^2 M_p^2 \left(\frac{128 \pi^2 M_p^2}{3 N_{\rm eff} (1+w)^2 H_{\rm inf}^2}\right)^{-\frac{3(w+1)}{3w-1}} \, , 
\end{equation}
which finally allows us to calculate the reheating temperature
\begin{eqnarray}\label{eq:TR}
& \frac{T_{R}}{ M_p} = \left( \frac{90}{\pi^2 g_*(T_R)} \right)^{1/4} \left( \frac{128 \pi^2}{3 N_{\rm eff} (1+w)^2} \right)^{-\frac{3(1+w)}{4(3w-1)}} \left( \frac{H_{\rm inf}}{ M_p} \right)^{\frac{3w+1}{3w-1}}.
\end{eqnarray}
We show the results as a function of the barotropic parameter $w$ in Fig~\ref{fig:Treh} for several values of $H_{\rm inf}$ and $N_{\rm eff}$. It is clear that for a given inflationary scale and $N_{\rm eff}$, reheating temperature grows with the barotropic parameter as this just corresponds to an inflaton redshifting away faster and leaving radiation to dominate at an earlier time. 
\\*

Radiation must dominate the Universe before $T_{\rm BBN}\approx 1 {\rm MeV}$ is reached to allow a successful nucleosynthesis~\cite{Cooke:2013cba,Patrignani:2016xqp}. Nevertheless, some contribution of the inflaton energy density at the MeV scale is still allowed. Let us define the Hubble rate $H$ and the usual radiation domination Hubble rate $H_r$
\begin{equation}\label{eq:HR}
H^2 = \frac{1}{3 M_p^2} (\rho_r + \rho_{\phi}) \, , \qquad H_r^2 = \frac{1}{3 M_p^2} \rho_r=\frac{1}{3 M_p^2}  \frac{\pi^2}{30}g_* T^4 \, .
\end{equation}
As discussed for instance in \cite{Lewicki:2016efe,Artymowski:2016tme} the upper bound on $H$ from BBN can be expressed as
\begin{equation}
\left. \left( \frac{H}{H_r} \right)^2 \right|_{T=T_{\rm BBN}} \leq 1 + \frac{7}{43} \Delta N_{\nu_{\rm eff}}\ \equiv \alpha \approx 1.038 \, ,
\end{equation}
where $\Delta N_{\nu_{\rm eff}} = 3.28 - 3.046$ is the difference between measured value and SM prediction for the effective number of neutrinos. 
This leads us to the relation
\begin{equation}\label{eq:alphabound}
\alpha-1 \geq \left. \frac{\rho_\phi }{\rho_r}\right|_{T=T_{\rm BBN}}=\left. \frac{\rho_\phi}{\rho_r}\right|_{\rm T=T_R} \left(\frac{a_{\rm BBN}}{a_R}\right)^{4-3(1+w)}=  \left(\frac{a_{\rm BBN}}{a_R}\right)^{1-3w} .
\end{equation}
We can approximate $\rho_{\rm BBN}$ as
\begin{equation}
\rho_{\rm BBN}\approx \rho_{r}(a_{\rm BBN})=\frac{\pi^2}{30} g_*\left(T_{\rm BBN}\right)T_{\rm BBN}^4 \, .
\end{equation}
Then using \eqref{ro_r} we find
\begin{equation}
a_{\rm BBN} =  a_{end} \frac{H_{\rm inf}}{\rho_{\rm BBN}^{1/4}} \left(\frac{9 N_{\rm eff} (1+w)^2}{128\pi^2}\right)^\frac{1}{4} \, , \ 
\end{equation}
which used in \eqref{eq:alphabound} together with \eqref{eq:aR} finally gives
\begin{equation}
 \frac{H_{\rm inf}}{M_p} \geq \left[ \left( \alpha - 1 \right)^{-\frac{1}{3w-1}} \frac{\left(\frac{1}{3} \rho_{\rm BBN} \right)^{1/4}}{M_p} \left( \frac{3 N_{\rm eff} (1+w)^2}{128 \pi^2} \right)^{-\frac{3}{4} \frac{1+w}{3w-1}} \right]^{\frac{3w-1}{3w+1}} \, . \label{eq:Hmin}
\end{equation}
This is the minimal $H_{\rm inf}$ required for the inflaton to redshift away sufficiently not to spoil BBN. Using \eqref{eq:TR} and \eqref{eq:Hmin} we find the minimal value of the reheating temperature
\begin{equation}
T_R \geq \left(\frac{30}{\pi ^2g_\star(T_R)}\right)^{1/4}\left(\alpha -1\right)^{-\frac{1}{3w-1}}
 \rho_{\rm BBN}^{\frac{1}{4}} \, ,
\end{equation}
which is $N_{\rm eff}$-independent. 
\\*

\begin{figure}[t]
\centering
\includegraphics[height=7.5cm]{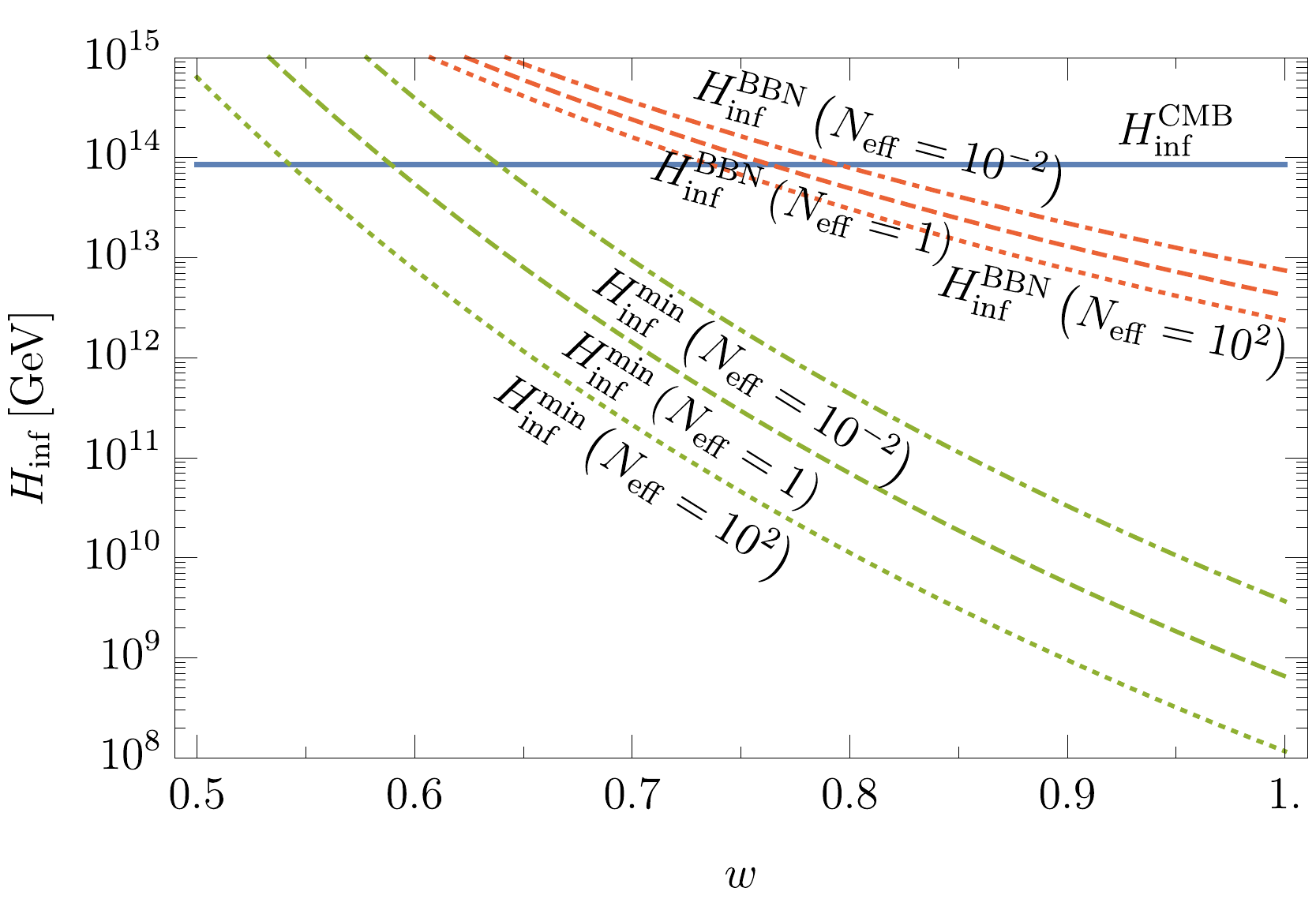}
\caption{\label{fig:Hminplot}
The minimal inflationary scale $H_{\rm inf}^{\rm min}$ allowing the energy density of the inflaton to redshift away enough not to spoil BBN constraints for $N_{\rm eff}=10^{-2}$, $N_{\rm eff}=1$ and $N_{\rm eff}=10^{2}$ (dot-dashed, dashed and dotted green lines respectively) together with the maximal $H_{\rm inf}^{\rm CMB}$ allowed by current CMB polarization data (solid blue line) and the maximal $H_{\rm inf}^{\rm BBN}$ allowed by the requirement not to spoil BBN by overproducing GWs for $N_{\rm eff}=10^{-2}$, $N_{\rm eff}=1$ and $N_{\rm eff}=10^2$ (dot-dashed, dashed and dotted red lines  respectively). The area between $H_{\rm inf}^{\rm CMB/BBN}$ and $H_{\rm inf}^{\rm min}$ is the allowed range of inflationary Hubble scales.
}
\end{figure}

\begin{figure}[t]
\centering
\includegraphics[height=7.5cm]{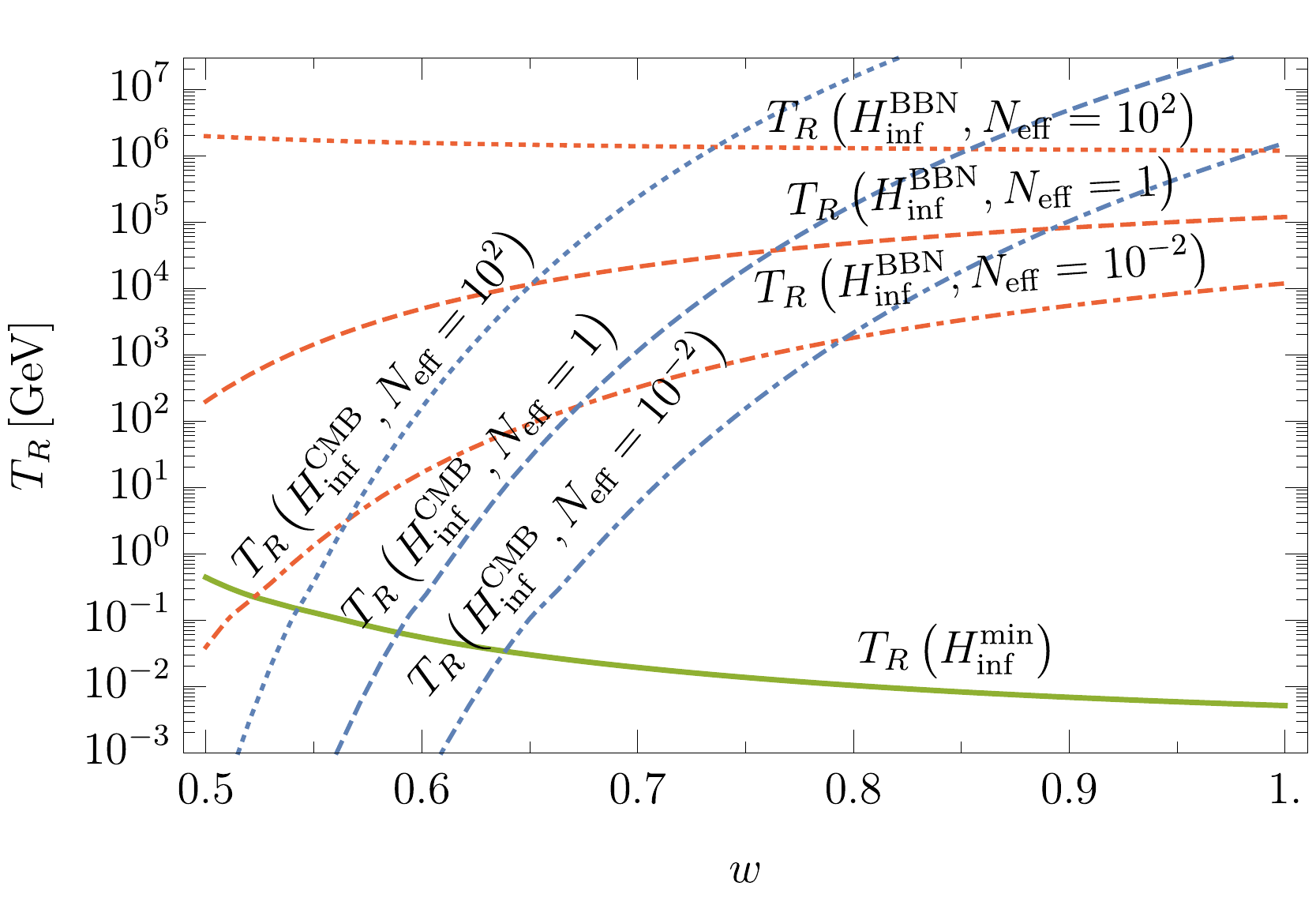}
\caption{\label{fig:Trehminplot}
The  minimal reheating temperature $T_R$ required for the inflaton to redshift away before BBN (solid green line) together with reheating temperatures corresponding to maximal inflationary scale allowed by CMB polarization data with $N_{\rm eff}=10^{-2}$, $N_{\rm eff}=1$  and $N_{\rm eff}=10^2$ (dot-dashed, dashed and dotted blue line) and maximal reheating temperature allowed by the requirement not to spoil BBN with overproduction of GWs for $N_{\rm eff}=10^{-2}$, $N_{\rm eff}=1$ and $N_{\rm eff}=10^2$ (dot-dashed, dashed and dotted red lines  respectively).  The area between $T_R(H_{\rm inf}^{\rm CMB/BBN})$ and $T_R(H_{\rm inf}^{\rm min})$ is the allowed range of reheating temperatures.
}
\end{figure}

The resulting minimal inflationary scale is simply such that radiation dominates slightly above the BBN temperature as shown in Figure~\ref{fig:Trehminplot}. As expected we can also see that the minimal reheating temperature rises for lower $w$ because it takes longer for the remaining inflaton energy density to redshift away sufficiently. It is also clear that low values of $w$ are already excluded because the minimal allowed $H_{\rm inf}$ is higher than the upper bound from the constraint from CMB polarization, which is $H_{\rm inf}\lesssim H_{\rm inf}^{\rm CMB}= 8.5\times 10^{13}$~\cite{Ade:2015xua}. Another constraint we include in Figure~\ref{fig:Hminplot} and Figure~\ref{fig:Trehminplot} $H_{\rm inf}\lesssim H_{\rm inf}^{\rm BBN}$ comes from the requirement not to overproduce gravitational waves (see Section~\ref{sec:GWs}) which would effectively act as extra radiation and spoil BBN predictions.
\\*


Note that for certain part of the parameter space one can still allow scales of inflation much smaller than the GUT scale (see Fig. \ref{fig:Hminplot}), which can be also useful in the context of the Higgs instability~\cite{Degrassi:2012ry,Buttazzo:2013uya}. Specifically if the scale of inflation is larger than the barrier between the electroweak vacuum and the true high energy minimum of the Higgs potential, that is $H_{\rm inf}\gtrsim 10^{10}-10^{12}$ GeV, the field would be pushed into the deeper minimum during inflation~\cite{Kobakhidze:2013tn,Hook:2014uia}. However, this reasoning assumes that SM is valid at least up to the instability scale and any modification below that scale can allow a higher inflationary scale.
\\*

\begin{figure}[ttt]
\centering
\includegraphics[height=7.2cm]{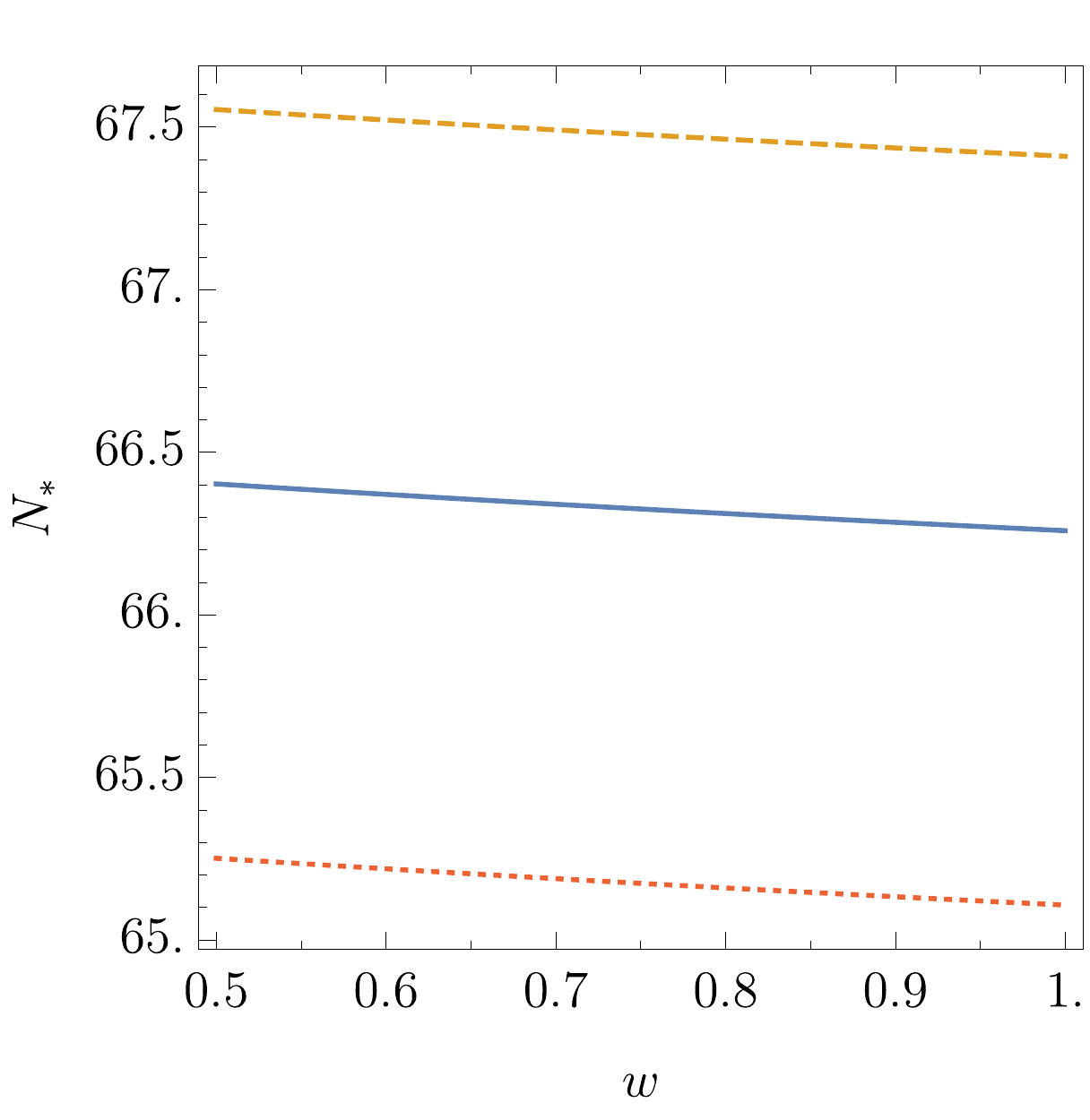}
\includegraphics[height=7.2cm]{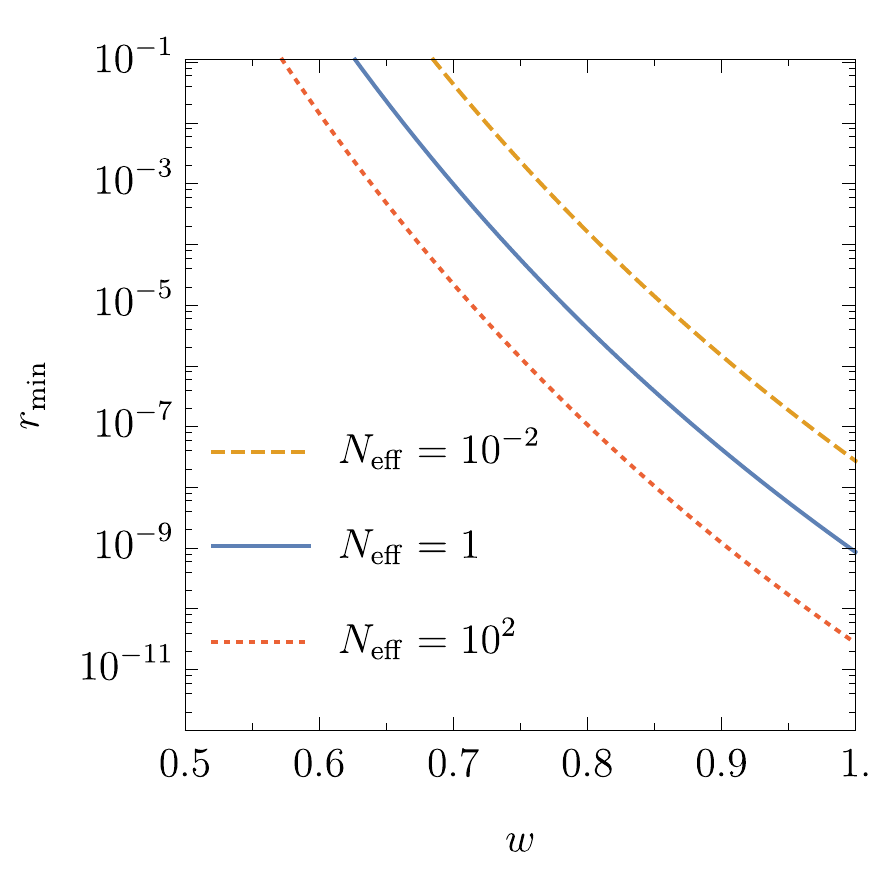}
 \caption{\label{fig:Nstar}
The values of $N_\star$ (left panel) and $r_{\rm min}$ (right panel) as a function of $w$ for several values on $N_{\rm eff}$ listed in the right panel. Note that most of specific realisations of the inflationary sector can be made consistent with these constraints from the tensor-to-scalar ratio.}
\end{figure}

In addition to the lower bound on $H_{\rm inf}$ we can also find the lower bound on $N_\star$. In Eq. \eqref{eq:Nstar} we use $V_{hor} \sim \rho_{end} \sim 3H_{\rm inf}^2$ and $\rho_{th} = \rho_{R}$, which for $k_{\star} = 0.002 \text{Mpc}^{-1}$ and for $\rho_r$ defined in \eqref{eq:rhoR} gives
\begin{eqnarray}
& N_\star \simeq 64.82 + \frac{1}{4} \ln\left( \frac{128 \pi^2}{ N_{\rm eff} (1+w)^2} \right). 
\end{eqnarray}
The result does not depend on $H_{\rm inf}$ and it significantly decreases the uncertainty of $N_\star$. This uncertainty mostly comes from the fact that the scales of inflation and reheating can both vary from the GUT scale to the MeV scale.
In dark inflation scenario there are only two free parameters which determine value of $N_\star$, namely $w$ and $N_{\rm eff}$. The barotropic parameter $w$ can take values larger than $1/3$ in order to provide sufficiently fast redshift of the inflaton after inflation. The number $N_{\rm eff}$ may vary more, from $N_{\rm eff}=1$ in the SM case with minimal coupling to gravity and $N_{\rm eff}\approx 100$ in the case of MSSM and in even grater range if the inflaton couples non-minimally to gravity. However, the dependence is only logarithmic and does not change $N_*$ dramatically as we show in the Fig.~\ref{fig:Nstar}.
\\*

We can also translate the bound on inflationary scale \eqref{eq:Hmin} to a constraint on the tensor-to-scalar ratio
\begin{eqnarray}
\frac{r}{0.01} \geqslant \left( \frac{M_p}{\Lambda_{\rm COBE}} \right)^2 \left[ \left( \alpha - 1 \right)^{-\frac{1}{3w-1}} \frac{ \left(\frac{1}{3} \rho_{\rm BBN} \right)^{1/4}}{M_p} \left( \frac{3 N_{\rm eff}(1+w)^2}{128 \pi^2} \right)^{-\frac{3}{4} \frac{1+w}{3w-1}} \right]^{2 \frac{3w-1}{3w+1}} 
\end{eqnarray}
using the the normalization condition 
\begin{equation}
\frac{r}{0.01} = \left( \frac{H_{\rm inf}}{\Lambda_{\rm COBE}} \right)^2 \, 
\end{equation}
with $\Lambda_{\rm COBE} = 2.54 \times 10^{13} \text{GeV}$.
The current upper bound $r < 0.09$, which sets the maximal scale of inflation as well, comes from the Planck/Bicep data \cite{Array:2015xqh}. The values of $r_{\rm min}$ as a function of $N$ and $w$ are plotted in the Fig. \ref{fig:Nstar}. It is clear that dark inflation can easily fulfill this limit as long as the value of $w$ is not too small. 
\\*

\section{Gravitational waves}\label{sec:GWs}
\begin{figure}[ttt]
\centering
 \includegraphics[height=6.5cm]{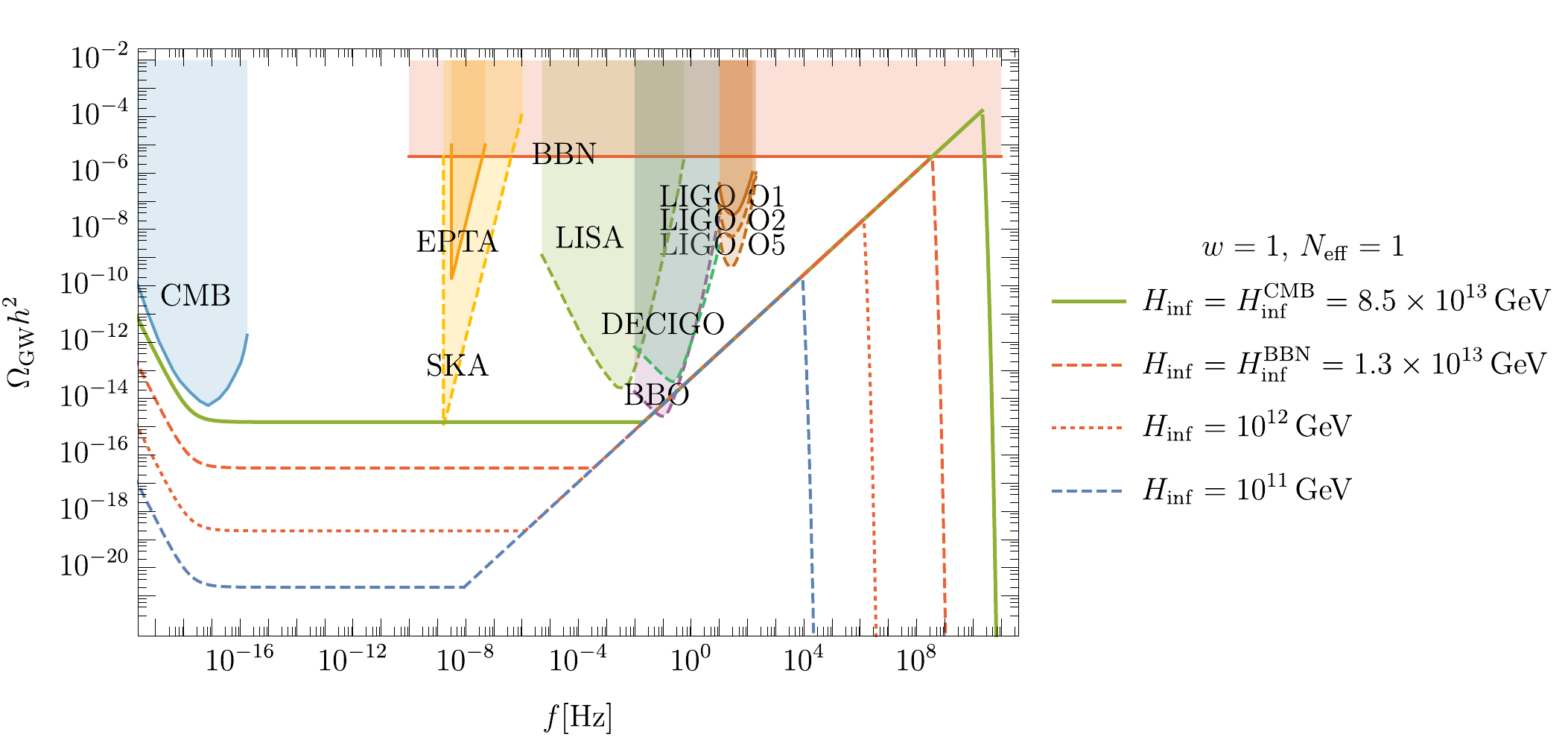}
 \caption{\label{fig:inflGW}
Gravitational wave energy density $\Omega_{\rm GW}h^2$ for several values of $H_{\rm inf}$ with $w=N_{\rm eff}=1$. We show the reach of current experiments using BBN data~\cite{Henrot-Versille:2014jua,Smith:2006nka} as well as present and planned LIGO capabilities~\cite{TheLIGOScientific:2014jea,TheLIGOScientific:2016wyq,Thrane:2013oya} and European Pulsar Timing Array~\cite{vanHaasteren:2011ni}. We also show the planned reach of future experiments LISA (with the most optimistic configuration A5M5) \cite{Bartolo:2016ami}, BBO and DECIGO \cite{Yagi:2011wg}, SKA~\cite{Janssen:2014dka} and CMB polarization \cite{Lasky:2015lej,Sepehri:2016jsu}.}
 \end{figure}

\begin{figure}[ttt]
\centering
 \includegraphics[height=6.5cm]{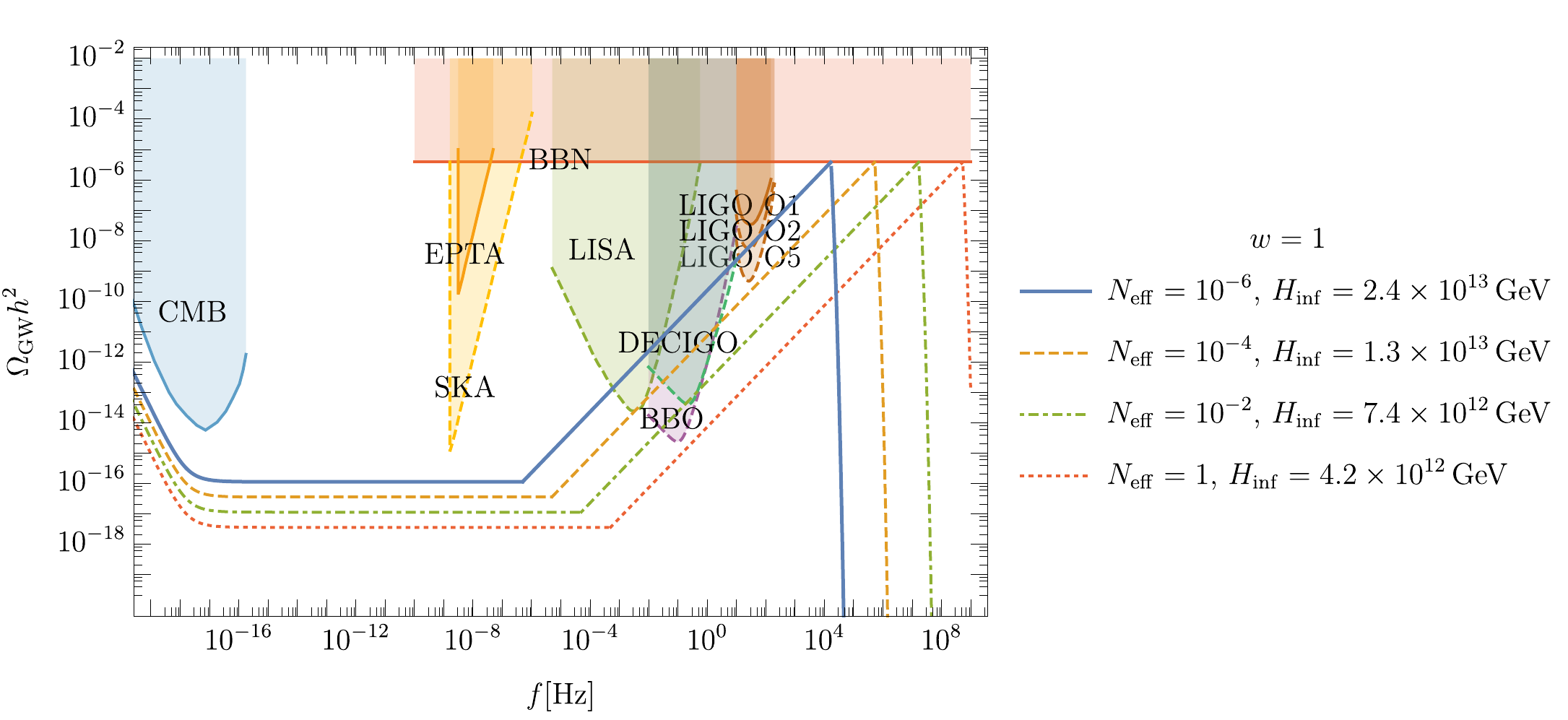} 
  \caption{
 \label{fig:inflGW2}
Gravitational wave energy density $\Omega_{\rm GW}h^2$ for several values of $N_{\rm eff}$ and maximal allowed $H_{\rm inf}$ in each case with $w=1$. The small values of $N_{\rm eff}$ may be easily obtained for inflation driven by a field $\phi$ non-minimally coupled to gravity $\xi\phi^2 R$ and for $\xi$ very close to the conformal value. In this case the signal from the inflationary gravitational waves could be observed in future experiments. Note that such spectra of primordial gravitational waves could in principle not only probe thermal history of the Universe, but also indicate a non-minimal coupling of the inflaton to gravity.}
\end{figure}

In this section we turn to the primordial gravitational wave spectrum from inflation. 
For simplicity we assume a scale invariant primordial power spectrum that can be approximated as~\cite{Langlois:2010xc}
\begin{equation}
P_{\rm GW}(k)= \frac{2 H_{\rm inf}^2}{\pi^2 M_p^2}.
\end{equation}
To compute the power spectrum today we need the transfer function~\cite{Turner:1993vb,Bartolo:2016ami} which allows us to describe the evolution of GWs in the late history of the Universe when matter domination begins. 
The transfer function is given by
\begin{equation}
T_{\rm T}^2(k,\tau_0)=\frac{3 \Omega_m j_1(k \tau_0)}{k \tau_0}\sqrt{1+1.36\left(\frac{k}{k_{\rm eq}}\right)+2.5\left(\frac{k}{k_{\rm eq}}\right)^2 },
\end{equation}
where $\Omega_m$ is the current matter abundance, $k_{\rm eq}$ is the scale of matter radiation equality and $j_1(x)\approx (\sqrt{2} x)^{-1}$ is the spherical Bessel function.
This late epoch is usually of interest because the most standard way of looking for inflation is through GWs in CMB polarization~\cite{Ade:2015lrj}. These waves re-enter the horizon at late times and very small scale which corresponds to a very low frequency $f\approx 10^{-17}-10^{-18} \, {\rm Hz}$.
This provides us with an upper bound on the inflation scale which currently is $H_{\rm inf}\lesssim 8.5\times 10^{-13}$.
 We will, however, be most interested in waves which re-enter the horizon much earlier when the inflaton field still dominates the expansion. The reason is that during this epoch GWs redshift much slower than the background which leads to an enhanced abundance today.
The GW spectrum today reads~\cite{Kuroyanagi:2014nba} 
\begin{equation}
\Omega_{\rm GW} h^2 (k,\tau_0)=,
 \begin{cases}
    \frac{k^2}{12 a_0^2 H_0^2} P_{\rm GW}(k) T_{\rm T}^2(k,\tau_0) & \ {\rm for} \ k \leq k_R \\
 \frac{k^2}{12 a_0^2 H_0^2} P_{\rm GW}(k) T_{\rm T}^2(k,\tau_0) \left(\frac{k}{k_R}\right)^{\frac{6w-2}{3w+1}} 
 & \ {\rm for} \ k_R<k \leq k_{ end}  ,
  \end{cases}
\end{equation}
where $k_R=a_R H(a_R)$ and during radiation domination it can be computed using Eq.~\eqref{eq:aR}-\eqref{eq:HR}. The highest reachable scale $k_{end}$ can be calculated using the same equations in the following way
\begin{equation}
\frac{k_{end}}{k_R}=\frac{a_{end}H(a_{end})}{a_{R}H(a_{R})}=\frac{1}{2}\left(\frac{\rho_{end}}{\rho_R}\right)^{\frac{1}{2}-\frac{1}{3(1+w)}}=\frac{1}{2}\left(\frac{128\pi^2 M_p^2}{3 N_{\rm eff} (1+w)^2 H_{\rm inf}^2}\right)^{\frac{1+3w}{2(1-3w)}}.
\end{equation} 
This increase of GW density at high frequencies gives hope for direct observation at planned experiments. As we show in Fig.~\ref{fig:inflGW} in the case of inflaton minimally coupled to gravity ($N_{\rm eff}\geq 1$) the modification occurs at frequencies slightly too high for direct observation in the near future. However, with non-minimal coupling close to its conformal value ($N_{\rm eff} \ll 1$) dark inflation could be probed by LIGO or future spaced based experiments in coming years as we show in Fig.~\ref{fig:inflGW2}. We focus our analysis on the $w=1$ scenario, since in this case the effect of the amplification of $\Omega_{\rm GW}$ is the strongest.

It is important to note that this behaviour of the GW spectrum at high frequencies depends on the evolution of the Universe  and is common for all scale invariant sources. Specifically a network of cosmic strings would result in an identical high frequency behaviour with a clear advantage coming from the fact that we could also observe the flat part of the spectrum and knowing the turning point compute the temperature at which non-standard cosmology takes over~\cite{Cui:2017ufi}. 

\section{Further consequences of dark inflation} \label{sec:dark} 

In this chapter we link dark inflation to existing models of dark energy and dark matter. It is important to note that all the previous results can be accommodated without modifications in realistic models explaining the cosmological evolution. Both cosmic inflation and present era of accelerated expansion may be caused by additional scalar fields or modified gravity with rather similar results. They both need to be characterized by the domination of the component with an equation of state close to $w \simeq -1$. Thus, it would not be unreasonable to try an explain both those eras of accelerated expansion with just a single underlying source. The simplest example here would be multi-plateau potential~\cite{Dimopoulos:2017zvq} with a GUT-scale plateau responsible for inflation and a dark energy plateau with energy scale of the order of $10^{-120}M_p^4$. Such an approach enables us to connect the present-day acceleration with the GUT scale physics, which could help to solve or at least weaken the hierarchy problem between the Planck scale and the dark energy scale. A model similar to \cite{Dimopoulos:2017zvq} was proposed in the context of multi-phase inflation~\cite{Artymowski:2016ikw}, where the multi-plateau potential was a result of embedding in a scalar-tensor theory. In principle this model could also be used as a source of both, dark inflation and dark energy, it would, however, induce a much more severe fine-tuning problem than in the case of Ref. \cite{Dimopoulos:2017zvq}. 
\\*

We assumed that the inflaton field is not coupled to Standard Model fields. Therefore the remains of the inflaton do not decay to SM fields and could conceivably play the role of dark matter. The case of a decoupled scalar DM candidate which in the early Universe evolves as a kinaton field ($w=1$) has already been discussed in the Refs. \cite{Rindler-Daller:2013zxa,Rindler-Daller:2015lua,Lewicki:2016efe}. The model consists of a scalar field $\phi$ with an action consisting of three terms - the canonical kinetic term $\dot{\phi}^2/2$, the mass term $m^2\phi^2/2$ and the quartic term $\lambda \phi^4$. Initially the kinetic term dominates, which is equivalent to taking the $w=1$ in our analysis. Therefore, one can consider the field to be a leftover from e.g. K-inflation, assuming that the potential terms where subdominant during inflation. After the $\dot{\phi}^2$ domination, the quartic term starts to dominate which for the oscillating scalar field is equivalent to $w=1/3$. Finally, the mass term comes to dominate and the field at late times behaves as a dark matter candidate.
\\*

Even without trying to directly connect DM and the inflaton, the non-standard evolution in dark inflation can have a dramatic effect on the DM relic abundance. This happens simply because the expansion rate is much bigger for $w>1/3$ than in the standard case which means that any standard particle DM candidate would drop out of thermal equilibrium earlier and remain with a higher abundance today.
Thus freeze-out of dark matter during inflaton domination would result in a much higher abundance~\cite{Beniwal:2017eik,Redmond:2017tja} \footnote{{This issue was also analysed in the Ref. \cite{Pallis:2005bb}}}.  

\section{Summary}

In this paper we investigated the cosmological consequences of inflation driven by a field, whose origin may be related to a modification of gravity, which couples to the Standard Model only gravitationally.
As a result such an inflaton does not decay after inflation and can dominate the energy budget of the Universe for an extended period after inflation.
In Sec. \ref{sec:reheating} we discussed gravitational reheating independently of the precise dynamics of the inflationary sector.
The reheating is dominated by purely gravitational particle production during the transition between a de-Sitter inflationary epoch and a post-inflationary epoch dominated by a perfect fluid with a barotropic parameter $w$.
In order to correctly return to the standard cosmological history at later times we need energy density of the remains of the inflaton to redshift faster than radiation, with $w>1/3$.
We determined the energy density of gravitationally produced radiation as well as the temperature of reheating, that is the temperature below which the usual radiation dominated cosmological history resumes.
We also calculated experimental bounds on the inflationary scale. The upper limit as usual comes from the bound on tensor-to-scalar ratio but there is also a competing bound coming from the BBN constraints on gravitational waves. This new bound originates in the requirement not to overproduce GWs during the inflaton domination. It becomes more severe for larger $w$ because the abundance of GWs is amplified during non-standard cosmological evolution.
The lower bound on the scale of inflation comes from the requirement that the remains of the inflaton field have to redshift away sufficiently so that they do not spoil the BBN.
\\*

We also show that the dark inflationary scenario implies significant consequences for the predictions of inflation. One of our main results is the calculation of $N_\star$, that is the number of e-folds before the end of inflation for which the pivot scale leaves the horizon, as a function of parameters of the model.
We argue that $N_\star$ does not depend strongly on the scale of inflation and on $w$, and for all our examples it is almost equal to $N_\star\simeq 66$. This sets a very precise constraint on inflation and defines the milestones of thermal history of the Universe with better accuracy. 
\\*

We also calculated the evolution of the primordial gravitational waves generated during inflation in Sec. \ref{sec:GWs}.
As we already mentioned the generic effect is an amplification of the signal during the period of inflaton domination.
We describe a generic situation but focus on the $w = 1$ case, which provides the strongest GW signal.
In most cases the signal is amplified at the frequencies too high to be observed in upcoming GW experiments.
A notable exception here is the case of $N_{\rm eff} \ll 1$ realised by non-minimal coupling to gravity close to its conformal value.
In this case the GW signal is amplified at low enough frequencies to be observed in the upcoming runs of LIGO and also planned detectors such as LISA, BBO or DECIGO. 
\\*

In the Sec. \ref{sec:dark} we discuss possible implementations of inflation in the dark sector and possible advantages of said implementations.
We point out that our scenario may be successfully realized in the model of Dimopoulos and Owen \cite{Dimopoulos:2017zvq}, in which a single scalar field is responsible for both inflation and dark energy domination. Another possibility is a model based our previous work \cite{Artymowski:2016ikw}.
In particular, it is interesting to note that the dark inflaton could also be the source of dark matter, for instance in a model of a scalar field with a transition between the kinaton and DM-domination phases. 
Finally, the influence of the "dark" thermal history of the Universe on the dark matter abundance can be non-negligible. The freeze-out of the DM particles could take place much earlier in the case of dark inflation, which can enhance  the DM relic density which in turn can significantly modify constrains on the DM models. 
\\*

To conclude, the dark inflation scenario may serve as a very attractive alternative to standard popular scenarios of reheating. It can be implemented in a broad class of inflationary models and can remain fully consistent with data. The scenario limits the uncertainty on $N_\star$ and it can produce a strong observable signal due to amplification of primordial gravitational waves.
\section*{Acknowledgements}

MA was supported by the Iuventus Plus grant No. 0290/IP3/2016/74 from the Polish Ministry of Science and Higher Education. OC was supported by the Polish NCN doctoral scholarship number 2016/20/T/ST2/00175. ZL was supported by the Polish NCN grant DEC-2012/04/A/ST2/00099. ML was supported by the Polish MNiSW grant IP2015 043174 and STFC grant number ST/L000326/1.

\bibliographystyle{JHEP}
\bibliography{inflation}

\end{document}